# Wave Propagation in Lipid Monolayers

**J. Griesbauer[1], A. Wixforth[1], M.F. Schneider[1*]**

[1] *University of Augsburg, Experimental Physics I, D-86159 Augsburg, Germany*

\* *Corresponding author:*

*Matthias F. Schneider*

*Current:*

*Phone: +49-821-5983311, Fax: +49-821-5983227*

*matschnei@gmail.dcom*

*University of Augsburg, Experimental Physics I,*

*Biological Physics Group*

*Universitätstr. 1*

*D-86159 Augsburg, Germany*

*New Address (September 1, 2009):*

*matschnei@gmail.dcom*

*Boston University, Dept. of Mechanical Engineering*

*110 Cummington St*

*Boston-Massachusetts, USA*

**Abstract**

Sound waves are excited on lipid monolayers using a set of planar electrodes aligned in parallel with the excitable medium. By measuring the frequency dependent change in the lateral pressure we are able to extract the sound velocity for the entire monolayer phase diagram. We demonstrate that this velocity can also be directly derived from the lipid monolayer compressibility and consequently displays a minimum in the phase transition regime. This minimum decreases from $v_0$=170m/s for one component lipid monolayers down to $v_m$=50m/s for lipid mixtures. No significant attenuation can be detected confirming an adiabatic phenomenon. Finally our data propose a relative lateral density oscillation of $\Delta\rho/\rho \sim$ 2% implying a change in *all* area dependent physical properties. Order of magnitude estimates from static couplings therefore predict propagating changes in surface potential of 1-50mV, 1 unit in pH (electrochemical potential) and 0.01°K in temperature and fall within the same order of magnitude as physical changes measured during nerve pulse propagation. These results therefore strongly support the idea of propagating adiabatic sound waves along nerves as first thoroughly described by Kaufmann in 1989 and recently by Heimburg and Jackson, but claimed by Wilke already in 1912.

## Introduction
The lipid monolayer is ubiquitously present in biology as one half of the cell, organelle or vesicle membrane. A complete understanding of the physical properties of lipid monolayers is therefore of fundamental interest in order to understand its role for biological bilayer systems. Studying sound propagation in insoluble organic films the lipid monolayer obtains its attraction from the fact that most physical properties (lateral pressure, area per molecule, compressibility, surface potential, temperature etc.) are easily accessible (1). Consequently lipid monolayers have been investigated very intensively from various viewpoints of physics and physical chemistry (see (1-4) and references therein). On the other hand only very little studies addressed internal excitations or wave propagation within lipid monolayers (5). If however two dimensional adiabatic excitations are present in simple lipid monolayer systems their absence in complex biological lipid membranes appears very unlikely.

Here, we present both experimental and theoretical studies on adiabatic sound wave propagation in lipid monolayers. Experimentally, lipid monolayers are excited by an alternating in-plane electric field (in-plane excitation electrodes, IPE) originating from a set of laterally patterned electrodes (interdigital transducers, IDT). For certain distinct frequencies $f$, we find pronounced changes in the lateral pressure which suggests electrical excitation with an underlying resonance phenomenon. This in turn allows to extract in detail the variable sound velocity of the propagating wave by matching the excitation frequency $f$ with the distance $d$ between two planar electrodes, defining an imposed wave length.

Thermodynamic analysis of the problem reveals that both the lateral pressure change as well as the sound velocity for the different thermodynamic states can be directly derived from the lipid monolayer phase diagram and gives clear evidence for the existence of a propagating, adiabatic sound wave. Order of magnitude estimates suggest that the origin of the decoupled propagation may be a consequence of the different sound velocities in 2D and 3D as well as the increased heat conductivity of the boundary water layer.

## Materials and Methods
Lipids 1,2-Dipalmitoyl-sn-Glycero-3-[Phospho-rac-(1-glycerol)] (DPPG) and 1,2-Dipalmitoyl-*sn*-Glycero-3-Phosphocholine (DPPC) dissolved in Chloroform were purchased from Avanti Polar Lipids (Birmingham, Al. USA) and used without further purification.

All measurements were done on a standard film balance with a heat bath (NIMA, Coventry, England) modified by a small stage to position the excitation chips with the IDT structure. Standard isotherms could be recorded and the regulation circuit of the film balance allowed it to hold a specific pressure of the monolayer for measuring the area expansion coefficient $\alpha$ (T).

The IPE chips were produced by standard lithography method, using LiTaO3 as substrate for the gold electrodes. The interdigitated gold electrodes (finger spacing w ≈ 10μm) were connected to an RF amplifier (ZHL-2010+, MINI-CIRCUITS, Brooklyn, NY, USA) which was driven by a signal generator (SML 01, Rhode&Schwarz, Munich, Germany). Such chips are routinely used in our lab for surface acoustic wave studies, including microfluidic and sensor applications. Here, however, we operate the chips to generate RF electric fields coupling to the lipid monolayers only and at frequencies where no surface waves of the substrate are resonantly excited.

To obtain lipid mixtures, lipids dissolved in chloroform were mixed in the desired proportions. Before spreading the lipid monolayers on the air water interface, the chip and the Wilhelmy plate were placed to the surface of the trough as described in Fig. 2. Frequency sweeps between f=0 and f=27 MHz were performed during 30 minutes, whereas the pressure was recorded simultaneously.

**Results and Discussion**

**Theory**

*Fundamental Thermodynamic Relations*

In Einstein's first publication in 1901 he approved our crucial assumption: applying a Carnot cycle to a free surface he demonstrated that the interface of a water drop must have its own heat (heat capacity) and therefore its own entropy $S^I$ (6, 7). Our experiments as well as order of magnitude estimates will demonstrate that a lipid monolayer at the air/water interface has its own entropy $S^I$ as well and has therefore to be considered as an independent thermodynamic system. To realize the meaning of $S^I$ for the experiment it is very helpful to recall that it is the second derivative of $S^I$ with respect to a thermodynamic variable $x$ that is related to the generalized susceptibilities by (8)

$$k_B \left( \frac{\partial^2 S^I}{\partial x_i^2} \right)^{-1} \equiv -k_B \left( \frac{\partial x_i}{\partial X_i} \right)_{X_{J \neq i}}. \qquad (1)$$

Here $k_B$ denotes the Boltzmann constant and $X_i = -\frac{\partial S^I}{\partial x_i}$ (e.g. $X_i = \frac{\pi}{T}, \frac{\mu}{T}, ...$) the corresponding thermodynamic force. For $x_i = A$ and constant temperature, this results in

$$k_B \left( \frac{\partial^2 S^I}{\partial A^2} \right)_T^{-1} = -k_B T A \kappa_T, \qquad (2)$$

where

$$\kappa_T = -\frac{1}{A} \frac{\partial A}{\partial \pi} \bigg|_T \qquad (3).$$

This is the lateral isothermal compressibility of the lipid monolayer, which can be directly derived from the pressure-area isotherm (Figure 1). Hence, the mechanical properties of the lipid monolayer represent a measure of the susceptibility, i.e. inverse curvature of the entropy potential. A maximum in $\kappa_T$ as observed in the phase transition regime corresponds to a "flat" entropy potential. It should be noted, that different boundary conditions (keeping $T$, $q$, $\mu$ or $N$, etc. constant), will result in different potentials like different 1D projections of the same n-dimensional potential.

*Estimation of $\kappa_{S^I}$ from $\kappa_T$*

In principle, the velocity $c_0$ could be estimated within an order of magnitude by using $\kappa_T$. The easy accessibility of the thermal expansion coefficient $\alpha$ however, enables us to get a better estimate of the adiabatic compressibility $\kappa_{S^I}$ using the thermodynamic relation (9)

$$\kappa_{S^I} \approx \kappa_T - \frac{T}{A c_\pi} \left( \frac{\partial A}{\partial T} \right)_\pi^2, \qquad (4)$$

where $c_\pi$ represents the heat capacity at constant pressure and $\alpha = \dfrac{1}{A}\left.\dfrac{\partial A}{\partial T}\right|_\pi$ the isobaric expansion coefficient. Even though this relationship is used quite commonly it should be noted that it strictly holds only for a system defined by $\pi$, $A$ and $T$ and will carry additional terms, when charge, dipole, chemical potential, etc. are included. Nevertheless, extracting $\kappa_T$ and $\alpha_\pi$ from the experiment and approximating $c_\pi$ by using the experimentally established correspondence between change in enthalpy $\Delta H$ and area $\Delta A$ (9)

$$\Delta H \approx \gamma \Delta A \tag{5a}$$

and therefore

$$\Delta c_\pi \approx \gamma \left(\dfrac{dA}{dT}\right)_\pi, \tag{5b}$$

Equation 4 can now be used to estimate the propagation velocity $c_0$ of a sound wave, which in the linear case is given from the fundamental thermodynamic relation

$$c_0 = \sqrt{1/\rho \kappa_{S'}}. \tag{6}$$

Here, $\rho$ is the area per molecule, an experimentally well controllable quantity. Since both $\kappa_T$ as well as $\kappa_T$ exhibit a maximum in the phase transition regime the sound velocity is expected to undergo a minimum near the isothermal phase transition of the lipid monolayer. Clearly, inseparable from the excited sound wave a temperature wave must propagate along the lipid monolayer as well. Knowing the propagating area density oscillation an estimate of the accompanied temperature change can be extracted from the isothermal expansion and the heat capacity of the monolayer by (10)

$$\Delta T = T c_0 v A \dfrac{\alpha_\pi}{c_\pi} \Delta \rho \approx T c_0 v \gamma^{-1} \Delta \rho, \tag{7}$$

where $v$ is the particle velocity and Eq. 5b has been applied as well. Again, even commonly used, this strictly holds only for a system defined by $\pi$, $A$ and $T$.

**Experiments**

**Excitation of Sound Waves by Planar Electrodes**

When a thermodynamic system is forced out of its equilibrium position, the conservation of entropy requires a propagation phenomenon. In this sense the observation of sound propagation in lipid monolayers would provide additional support that the lipid membrane interface can be considered as a "closed" two dimensional system, fairly well decoupled from the surrounding bulk bath. Qualitatively a flat thermodynamic potential implies weak restoring forces and therefore a lower propagation velocity as for steep potentials.
In our experiments the excitation of propagating sound waves along the lipid membrane was accomplished by incorporating a chip with a planar array of gold electrodes (in-plane excitation, IPE) in the plane of a negatively charged DPPG monolayer (Figure 2). Traditionally, such IDTs are used as filters in RF applications or to create acoustic streaming in microfluidic systems (11, 12). Another application of IDT are sensors comparable to the well known quartz crystal microbalance (QCM)(13, 14). Here, however, we use such electrodes to excite in-plane waves on soft interfaces at various frequencies by electro-mechanical coupling to a polarisable membrane. We employ the fact that a lateral density oscillation $\rho = \rho_0 + B \cdot \cos \omega t$ creates a net pressure increase $\Delta \pi$. This is due to a nonlinear pressure-area relationship $\pi(A)$ and can be calculated with the help of the isotherm by simply integrating the pressure change over one period:

$$\overline{\Delta \pi} = \frac{\omega}{2\pi} \int_0^{\frac{2\pi}{\omega}} \pi(A(t))dt - \pi(A) \qquad (8a).$$

The same expression can be written in terms of the mass density $\rho$:

$$\overline{\Delta \pi} = \frac{\omega}{2\pi} \int_0^{\frac{2\pi}{\omega}} \pi\left(\frac{m_{Lipids}}{\rho_0 + B \cdot \cos \omega t}\right)dt - \pi\left(\frac{m_{Lipids}}{\rho_0}\right) \qquad (8b).$$

Here, $\omega$ is the angular frequency of the travelling wave, $B$ its amplitude, $\rho_0$ the lateral density in rest and $m_{Lipids}$ the mass of the lipids forming the monolayer. The first term on the right side of Eq. 8 represents the time averaged pressure change due to the modulation in the interface, while the second term denotes the undisturbed lipid monolayer.

If adiabatic propagation would indeed take place, Eq. 8 predicts a net increase in lateral pressure. The pressure spectrum of a DPPG monolayer shown in Fig. 3 confirms that such an increase takes place indeed. The average change in lateral pressure $\overline{\Delta \pi}$ detected during the excitation of the wave is plotted as a function of the stimulating frequency applied to the chip for three different configurations. Between f = 100 kHz and f = 27 MHz a pure water surface does not produce any significant response. Nevertheless at a distance of 2cm at around f=33 MHz pure water may also show a response. The origin of this signal is presently still unknown but may be attributed to surface capillary water waves (15). The presence of a DPPG monolayer on this surface, however, results in a pronounced variations in $\overline{\Delta \pi}$ around f = 11 MHz. Importantly the response at f = 11MHz is also not visible on the pure water surface at shorter distances between excitation and detection. We therefore conclude, that the change of $\overline{\Delta \pi}$ around f = 11MHz can ubiquitously be attributed to the presence of the lipid membrane and will be used for further data interpretation.

**Attenuation and Amplitude of the Wave**

Considering the macroscopic (15cm) distance between excitation and detection, the propagation of the wave does not seem to be significantly attenuated. To experimentally verify this finding, we measured the change in lateral pressure $\overline{\Delta \pi}$ as a function of distance from the source of excitation (see inset of Fig. 3). It turns out that the experimental data can be well fitted by a polynomial $\overline{\Delta \pi}(x) \propto -(x-b)^{-2}$, but not with an exponential decay function. This supports the idea of an adiabatic wave with decay due only to the geometry of the system but little dissipation.

Finally, the amplitude of the excited wave can be estimated by comparing the experimentally observed $\overline{\Delta \pi}$ to Eq. 8 and reveals a density variation B ≈ $0.02\rho_0$ (± $0.005\rho_0$). This in turn would result in a density modulation amplitude of about B ~0,3 ·$10^{-7}$kg/$m^2$, which is well within reasons for lipid monolayers.

**Sound velocity from lateral pressure changes**

The observed $\overline{\Delta\pi}$ clearly exhibits a frequency dependence (Fig. 3) with an apparent resonance like feature at f = 11MHz, indicating significant excitation at this point. It appears, that sound velocity $c_0$, electrode spacing $d$ and stimulating frequency $v_0$ are ideally matched at exactly this frequency in order to provide effective excitation. Taking the finger spacing between two electrodes $d = 12 \mu m$ (therefore $\lambda$ = 24 µm) and the resonance frequency $v_0$ = 11MHz from the experiment, we find a propagation velocity of $c_0=\lambda v_0$ = 260m/s.
Following the same procedure, we repeated the experiment along the entire isotherm also including the phase transition regime and extracted the corresponding sound velocity. It turns out that the different thermodynamic states of the monolayer indeed exhibit different excitation frequencies $v_0$ as can be seen in Fig. 4 for three different surface pressures.
A critical test whether the origin of the observed pressure change is indeed a lateral density oscillation arises from the fact that Eq. 8 predicts a negative change in $\overline{\Delta\pi}$ in and below the maximal compressibility, while $\overline{\Delta\pi}$ is positive for pressures above the maximum. Clearly, this behaviour is qualitatively and quantitatively reproduced in Figure 4 and does therefore strongly support our assumption of a propagating sound wave along the interface.

**Sound velocity from the monolayer compressibility**

The experiments described above enabled us to extract the sound velocities along the entire isotherm. If this represents indeed a two dimensional sound wave the velocity $c_0$ should directly depend on the adiabatic lateral compressibility (Eq. 6). Even though $\kappa_T$ will produce the right order of magnitude for the sound velocity, we exploit the fact that all other susceptibilities appearing in Eq. (4) can be extracted from the monolayer isotherm as well and will provide the basis to calculate a more accurate $\kappa_S$. Therefore $\alpha_\pi$ was measured from the $A(T)$ isobars for different lateral pressures, to calculate a more accurate approximation of $\kappa_S$. $c_\pi$ was calculated by applying Eq. 5, which should give a good estimate at least close to the phase transition regime (9). Finally, comparing the propagation velocities as being calculated from Eq. 6 and the measured sound velocities using $c_0=\lambda v_0$, we find both qualitative and quantitative agreement within 10% or less (Fig. 5a) confirming the existence of a propagating sound wave.
In particular, the minimum in propagation speed as a consequence of the maximum in $\kappa_T$ is resolved. The same qualitative behaviour is observed for a mixture of DPPG/DPPC (1:10). Again excellent quantitative agreement is achieved between $c_0$ calculated from Eq.4 and the experimental pressure spectra (Fig. 5b). For this lipid, however, the minimum velocity predicted from the isotherm and confirmed experimentally is only $c_0$ = 50m/s corresponding to the phase transition regime between liquid expanded and liquid condensed phase. This finding illustrates that both physical parameters (lateral pressure, temperature etc.) and the composition of the monolayer control the propagation velocity.

**On the coupling between monolayer and bulk**

The only reasonable propagation mechanism for our system is one in which heat and entropy $S_I$ (Eq. 1-3) of the interface are approximately conserved (no exponential decay in wave energy). Therefore, the fact that only *weakly* damped wave propagation is experimentally observed (see inlet of Fig. 3) calls for a decoupling between monolayer and bulk. Although, this is an experimental result and strongly supported by Einstein's early work on the heat of surfaces (7), we would like to outline another argument that assumes coupling but will demonstrate its insignificance at the same time. Following closely the book of Landau & Lifshitz (Vol 6) chapter II, V and VIII (10), we present an order of magnitude estimate of why the lipid monolayer may be decoupled from the bulk.

When a wave propagates in media 1 (monolayer) at the interface to an adjacent media 2 (bulk water), the reflection coefficient $R$ depends on the angle of incident $\theta$, the density $\rho$ and sound velocity $c$ of the two media (10, VIII).

$$R = \left[ \frac{\rho_2 c_2 \cos\theta - \rho_1 \sqrt{(c_1^2 - c_2^2 \sin^2\theta)}}{\rho_2 c_2 \cos\theta + \rho_1 \sqrt{(c_1^2 - c_2^2 \sin^2\theta)}} \right]^2 \qquad (9)$$

If the incident wave forms an angle of less than the critical angle $\theta_c$, where $\sin\theta_c = c_1/c_2$ (10, VIII), the entire wave is reflected (total internal reflection). Taking $c_1 \sim 100$ m/s and $c_2 \sim 1500$ m/s we obtain a critical angle of $\theta_0 \approx 5°$ for our arrangement. Lateral waves excited within the lipid monolayer will therefore be completely reflected. In this situation Landau (10, II) demonstrates that oscillations parallel to the interface of exponential decay in amplitude are created. These oscillations, however, do not provide significant dissipation as their exponential penetration into the bulk only resembles the energy distribution of the oscillating, propagating sound wave around the interface and must not be mistaken with the actual transport of energy out of the system, but as *inseparable* from the wave.

Similar arguments hold for the dissipation of heat. If the heat remains within the typical spatial extension of the sound wave, dissipation cannot be significant. As the temperature variations in the system are a direct consequence of the wave oscillations the temperature changes perpendicular to the interface must be exponential as well (10, V)

$$T = T_0 \exp(-z\sqrt{\frac{\omega}{2\chi}}) \exp(i\left[z\sqrt{\frac{\omega}{2\chi}} - \omega t\right]) \qquad , \qquad (10a)$$

where $\omega = 2\pi f$, and $\chi$ is the thermometric conductivity which can be calculated from the heat capacity $c_P$, the density $\rho$ and the thermal conductivity $k$ of the system (10, V)

$$\chi = \frac{k}{c_P \cdot \rho}. \qquad (10b)$$

Following Landau (10, II) the extension or viscous penetration depth $\delta$ in z-direction (Fig. 6) can be estimated from

$$\delta = \sqrt{\frac{2\eta}{\rho\omega}} \qquad , \qquad (11)$$

where $\eta$ is the dynamic and $\eta/\rho$ the kinematic viscosity of water. Using standard numbers for water, this depth turns out to be $\delta \approx 300$nm. In order to estimate whether thermal diffusion can add significantly to the dissipation process we need to compare $\delta$ with the thermal penetration length $\xi$

$$\xi = \sqrt{\frac{2\chi}{\omega}} \quad , \tag{12}$$

where $\chi$ denotes the thermometric conductivity (see Eq. 10b). This length scale describes over which distance a significant change in temperature takes place. Using $c_p \approx 4$ kJ(kgK)$^{-1}$ and $k \approx$ 0.6 J(mKs)$^{-1}$ we arrive at $\xi \approx 100$nm. Since $\delta \geq \xi$ the heat is, even when assuming coupling to the bulk, unable to "escape" the spatial extension of the sound wave within the timescale of compression ($0.5 \cdot 10^{-7}$s here). In other words Eq. 10a simply describes the *reversible* temperature oscillations inside the sound wave.

Finally, heat could also dissipate *within* the monolayer plane. This is to be predicted from a two dimensional shear viscosity observed in lipid monolayers. To estimate the corresponding loss inside the monolayer, we may also compare the length scale $l$ of heat dissipation within the time of one compression/expansion cycle $t = f_0^{-1}$ to the wavelength $\lambda$. Only if the heat expands (diffuses) during $t$ over a distance $l$, which is of the same order of magnitude as the wavelength $\lambda$, significant dissipation is to be expected. According to Landau, $l$ can be calculated from the solution of the general equation of heat transfer (10, V):

$$l^2 \approx \chi\tau \quad , \tag{13}$$

where $\chi$ is the aforementioned thermometric conductivity (Eq. 10b). Unfortunately, to our knowledge, no numbers for the heat conductivity $k_I$ of lipid monolayers exist. However, $k_I$ may be approximated by the heat conductivity of interfacial or boundary water, which has experimentally been observed on lipid membranes (16, 17). Using $k_I \approx 6$ *J/mKs* (which is one order of magnitude larger than for bulk water) (18), $c_p \approx 10$ *kJ/kgK* (19), and $\tau \approx t$ the lateral extension of heat due to thermal diffusion $l \approx 200$nm, which is ~ 100 times smaller than the wavelength ($\lambda \sim 24\mu$m). Therefore, the heat cannot "escape" the propagating wave in the lateral direction either.

To summarize this paragraph, we would like to state that neither mechanical nor thermal coupling properties support the admittedly intuitive prejudice of a strongly attenuated (since coupled) wave, but are in agreement with our experimental observation of a propagating, only weakly attenuated sound wave.

**Conclusion and Biological Impact**

A new approach to excite and detect acoustic waves in lipid monolayers is presented. Moreover, the existence of adiabatic sound waves is experimentally confirmed and the corresponding sound velocities are extracted from our measurements. Comparison of our findings with theoretical predictions provide excellent agreement and reveal velocities between 300 m/s and 170 m/s for one component, and 300 m/s to 50 m/s for two component systems, with a distinct minimum in the phase transition regime. Considering the simplicity of our system, these values are in very good agreement with reported propagation velocities of action potentials in nerves, ranging between 10 – 100 m/s and depending not only on myelination but also on temperature, thickness, sodium concentration etc. (*20-23*).

Our results provide an explanation of the well known, yet still unresolved problem of a temperature variation which is observed during action potentials (24-27). Our findings even predict that such reversible changes must occur. Quantitative estimates in *ΔT(t)* calculated from static experiments and the modulation in area density $\rho(t)$ using Eq. 7, propose temperature variations in the lipid monolayer of 0.01°K. Propagating changes in surface potential are expected as well. Taking surface potential measurements from static, isothermal experiments (2), the observed variation in area density $\rho(t)$ predicts a change in $\Delta U$ between 1 mV and 50 mV propagating along the surface of the lipid monolayer. Our results are therefore in support of the idea of propagating sound waves in biological membranes as first discussed by Wilke (28) and Wilke and Atzler (29) in 1912, first thoroughly described by Kaufmann in 1989 (30, 31), and recently discussed by Heimburg and Jackson in 2005 (24). The reported changes in temperature and pressure observed during nerve-pulse propagation (24-27) are at least in qualitative agreement with our predictions.

**Acknowledgement**
We thank Dr. K. Kaufmann (Göttingen) for very helpful discussions and highly recommend the reader to consult his earlier work [s Ref. 30 and 31]. MFS likes to personally thank K. Kaufmann who inspired him to work in this field and introduced him to the thermodynamic origin of propagation along membranes and nerves.
Financial support by the Bundesministerium für Bildung und Forschung is gratefully acknowledged. MFS likes to thank the Bavarian Science Foundation for financial support. This work has also been partially funded by the German Excellence Initative 'NIM'.

**References**

1.  Gaines, G. L. 1966. Insoluble Monolayers at Lipid-Gas Interfaces. *Interscience, New York.*
2.  Möhwald, H. 1995. Structure and Dynamics of Membranes. *Elsevier: Amsterdam.*
3.  Albrecht, O., H. Gruler, and E. Sackmann. 1978. Polymorphism of phospholipid monolayers. *Journal de Physique I 39:301-313.*
4.  McConnell, H. M. 1991. Structures and Transitions in Lipid Monolayer at the Air-Water Interface. *Annu Rev Phys Chem 42:171 - 195.*
5.  Frey, W., and E. Sackmann. 1992. Solitary waves in asymmetric soap films. *Langmuir 8:3150-3154.*
6.  Einstein, A. 1901. Folgerungen aus den Capillaritätserscheinungen. *Ann d Phys 4:513-523.*
7.  Einstein, A. 1910. Theorie der Opaleszenz von homogenen Flüssigkeiten und Flüssigkeitsgemischen in der Nähe des kritischen Zustandes. *Ann d Phys 25:205-226.*
8.  Landau, L. D., and E. M. Lifschitz. 1987. Course of Theoretical Physics Vol. 5. *Butterworth-Heinemann.*
9.  Heimburg, T. 1998. Mechanical aspects of membrane thermodynamics. Estimation of the mechanical properties of lipid membranes close to the chain melting transition from calorimetry. *Biochim Biophys Acta 1415:147-162.*
10. Landau, L. D., and E. M. Lifschitz. 1987. Course of Theoretical Physics Vol. 6. *Butterworth-Heinemann.*
11. Schneider, M. F., S. W. Schneider, V. M. Myles, U. Pamucki, Z. Guttenberg, K. Sritharan, and A. Wixforth. 2007. An Acoustically Driven Microliter Flow Chamber on a Chip (µFCC) for Cell/Cell and Cell/Surface Interaction Studies. *ChemPhysChem. 9:651-655.*

12. **Schneider, S. W., S. Nuschele, A. Wixforth, C. Gorzelanny, A. Alexander-Katz, R. R. Netz, and M. F. Schneider. 2007. Shear-induced unfolding triggers adhesion of von Willebrand factor fibers.** *Proc Natl Acad Sci U S A 104:7899-7903.*
13. **Janshoff, A., H. J. Galla, and C. Steinem. 2000. Piezoelectric Mass-Sensing Devices as Biosensors-An Alternative to Optical Biosensors?** *Angew Chem Int Ed Engl 39:4004-4032.*
14. **Freudenberg, J., M. von Schickfus, and S. Hunklinger. 2001. A SAW immunosensor for operation in liquid using a SiO/sub 2/ protective layer.** *Elsevier. Sensors & Actuators B Chemical:1-3.*
15. **Il'ichev, A. 1998. Self-Channeling of surface water waves in the presence of an additional surface pressure.** *Eur J Mech B 18:501-510*
16. **K. Åman, E. Lindahl, O. Edholm, P. Håkansson, P. Westlund. 2003. Structure and Dynamics of Interfacial Water in an Lα Phase Lipid Bilayer from Molecular Dynamics Simulations.** *Biophysical Journal, Vol. 84, 102-115.*
17. **V.Luzzati, F. Husson. 1962. The Structure Of The Liquid-Crystalline Phases Of Lipid-Water Systems.** *J. Cell. Biol., Vol. 12, 207-219.*
18. **J. S. Clegg, W. Drost-Hansen. 1991. On the biochemistry and cell physiology of water.** *P. W. Hochachka and T. P. Mommsen, 1–23.*
19. **A. Blume. 1983. Apparent Molar Heat Capacities of Phospholipids in Aqueous Dispersion. Effects of Chain Length and Head Group Structure.** *Biochem., Vol. 22 (23), 5436–5442.*
20. **A. F. Huxley, R. Stampfli. 1949. Evidence for saltatory conduction in peripheral myelinated nerve fibres.** *J. Physiol., Vol. 108, 315-339.*
21. **H. P. Ludin, F. Beyeler. 1977. Temperature dependence of normal sensory nerve action potentials.** *J. Neurol., Vol. 216(3), 173–180.*
22. **J. B. Hursh. 1939. Conduction Velocity and Diameter of Nerve Fibres.** *Am. J. Physiol., Vol. 127, 131-139.*
23. **W. L. Hardy. 1973. Propagation Speed in Myelinated Nerve.** *Biophys. J., Vol. 13, 1054-1070.*
24. **Heimburg, T., and A. D. Jackson. 2005. On soliton propagation in biomembranes and nerves.** *Proc Natl Acad Sci U S A. 102:9790-9795.*
25. **Ritchie, J. M., and R. D. Keynes. 1985. The production and absorption of heat associated with electrical activity in nerve and electric organ.** *Q Rev Biophys 18:451-476.*
26. **Tasaki, I. 1995. Mechanical and thermal changes in the Torpedo electric organ associated with its postsynaptic potentials.** *Biochem Biophys Res Commun 215:654-658.*
27. **Kim, G. H., P. Kosterin, A. L. Obaid, and B. M. Salzberg. 2007. A Mechanical Spike Accompanies the Action Potential in Mammalian Nerve Terminals.** *Biophys J.*
28. **Wilke, E. 1912. Das Problem der Reizleitung im Nerven vom Standpunkt der Wellenlehre aus betrachtet.** *Pflügers Arch 144:35-38.*
29. **Wilke, E., and E. Atzler. 1912. Experimentelle Beiträtge zum Problem tier Reizleitung im Nervem.** *Pflügers Arch 146:430-446.*
30. **Kaufmann, K. 1989. On the role of the phospholipid membrane in free energy coupling.** *Caruaru, Brazil (http://membranes.nbi.dk/ Kaufmann/pdf/ Kaufmann book5 ed.pdf).*
31. **Kaufmann, K. 1989. Action Potentials and Electrochemical Coupling in the Macroscopic Chiral Phospholipid Membrane.** *Caruara, Brazil (http://membranes.nbi.dk/ Kaufmann/pdf/ Kaufmann book4 ed.pdf).*

**Figure Captions**

**Fig. 1** Isotherms of lipid monolayers for 1,2-Dipalmitoyl-sn-Glycero-3-[Phospho-rac-(1-glycerol)] (DPPG) at 14°C and for a mixture of 1,2-Dipalmitoyl-*sn*-Glycero-3-Phosphocholine (DPPC) to DPPG as 10:1 at 20°C. Both Isotherms show maxima in the compressibility at about 8 mN/m for DPPG and at about 2,5 mN/m for the mixture, whereas the later clearly corresponds to the phase transition regime. The inset shows the resulting compressibility of the corresponding monolayer. As for the in-plane excitation (IPE) setup, a standard film balance was used for the measurements.

**Fig. 2** The in-plane excitation (IPE) setup. A set of electrodes, (IDTs), is used to create a local (electrode distance ≈ 10μm) oscillating electric field in the plane of the lipid monolayer, which excites a propagating wave by local lateral polarization of the membrane. The substrate is $LiTaO_3$ being coated with $SiO_2$. To establish a tight contact between the IPE chip and the lipid monolayer for the detection, the $SiO_2$ is additionally rendered hydrophobic (silanized). The change in lateral pressure is determined by the force on a Wilhelmy plate dipped into the monolayer.

**Fig. 3** Frequency spectra of the nonlinear pressure change induced by the IPE chip in a DPPG lipid monolayer (red curve) or pure water for different distances d (black for d=15cm (solid line) and d=2cm (dotted line)). At lower frequencies, the applied electric field does not affect the average lateral pressure, being measured by a Wilhelmy plate (see Fig. 1). However, at certain frequencies (here 10,9MHz and 33MHz), the lateral pressure significantly drops. The critical frequency ($v_{crit}$) on the x-axis can be changed by changing the overall lateral pressure. We calculate the propagation velocity ($c_0 = \lambda v_{crit}$) from the finger spacing of the stimulating electrodes and the measured critical frequency $v_{crit}$. The inset shows that attenuation is negligible. In fact, an exponential fit would lead to a decay length of ~1m.

**Fig. 4** Frequency spectra of the lateral pressure change, measured for different surface pressures in a DPPG lipid monolayer at T=14°C. The pressures are chosen to represent the different phases of the monolayer. The critical frequencies, were the nonlinear pressure change occurs for the first time, are marked. During maximal compressibility, the critical frequency is at a clearly lower value, than for the other phase states. Note, that $\overline{\Delta \pi} > 0$ above the maximal compressibility and $\overline{\Delta \pi} < 0$ elsewhere. This is correctly predicted by the assumption of an oscillating density wave (Eq. 8).

**Fig. 5** Propagation velocities along the lipid monolayer, measured as described in Fig. 2 [filled squares] and calculated from the adiabatic compressibility as described in the text [line]. The inserts show the experimentally obtained isothermal compressibility used for the calculation of $c_0$. (a) shows the velocity curves for DPPG and (b) for a mixture of DPPG and DPPC (1:10). For high lateral pressures, the agreement between $c_0$ as being predicted from the adiabatic compressibility and the experimentally determined velocity is within less than 5%. In the phase transition and for the lower pressure regime the deviation increases, but still falls within a 10% error bar. The origin of the deviation may be due to the fact that the monolayer becomes less stable at low pressures. Note the larger errors around the maximum in compressibility. These probably correspond to the increased sensitivity of $\kappa_s$ in to oscillations in density in that region.

**Fig. 6** Schematic representation of the important length scales of the system. Both viscous ($\delta$) and thermal ($\xi$) penetration depths and the geometry of the system are marked. While the lateral density wave propagates along the x-y-plane, oscillations both in density and in temperature decay exponentially along the z-direction. The density length scale ($\delta$) is of the same order or larger than the thermal penetration depth ($\xi)$ and therefore proposes negligible dissipation of heat into the bulk water.